\documentclass[aps,prd,groupedaddress,showpacs,showkeys]{revtex4}
\usepackage{amssymb}

\begin{document}

\title{Decays of tensor mesons and the tensor glueball in an effective field
approach}
\author{F. Giacosa, Th. Gutsche, V. E. Lyubovitskij and Amand Faessler}
\affiliation{Institut f\"ur Theoretische Physik, Universit\"at T\"ubingen, \\
Auf der Morgenstelle 14,D-72076 T\"ubingen, Germany\\
}
\date{\today}

\begin{abstract}
The strong and electromagnetic decays of the ground-state tensor mesons are
studied in an effective field approach. A fit to the well-known experimental
data is performed. The decay ratios of the tensor glueball are evaluated and
possible candidates are discussed.
\end{abstract}

\pacs{12.39.Fe, 12.39.Mk, 13.25.Jx, 13.40.Hq}
\keywords{Tensor mesons, glueball, effective chiral approach, strong and
electromagnetic decays}
\maketitle

\section{Introduction}

The experimental state of knowledge about the tensor meson nonet $2^{++}$ is
well established: the identified resonances $f_{2}(1270),$ $f_{2}^{\prime
}(1525),$ $a_{2}(1320)$ and $K_{2}^{\ast }(1430)$ are listed by the PDG~\cite%
{Eidelman:2004wy} as the lightest tensor states and correspond to the $2^{++}
$ ground state nonet (two isoscalars, an isotriplet and two isodoublets,
respectively). We do not consider here the resonances $f_{2}(1430)$ and $%
f_{2}(1565),$ omitted from the summary table of~\cite{Eidelman:2004wy}.
Strong decays of tensor states into two-body modes have been measured by
various experiments and the corresponding averages (or fits) reported in
Ref.~\cite{Eidelman:2004wy} are precise and well determined. Also the
two-photon decays of the tensor states $T\rightarrow \gamma \gamma$ are
well-known. Among the p-wave quark-antiquark nonets ($0^{++},1^{++},2^{++}$)
the tensor mesons are the ones which are experimentally best analyzed.
Tensor mesons have been studied using different theoretical methods:
effective Lagrangian approaches based on vector- and tensor-meson dominance~%
\cite{Renner:1971sj}-\cite{Cotanch:2005ja}, the naive quark model with
possible meson-glueball mixing~\cite{Lee:1978py}-\cite{Carvalho:2000ir},
current-algebra approach~\cite{Lahiri:1980rk}, lattice QCD~\cite%
{Teper:1987ws}, QCD sum rules~\cite{Bagan:1988ay}, the $^3P_0$ model~\cite%
{Gobbi:1993au,Ackleh:1996yt}, Chiral Perturbation Theory (ChPT)~\cite%
{Bellucci:1994eb}-\cite{Chow:1997sg}, Regge model~\cite{Anisovich:2004vj},
dispersion-relation technique~\cite{Surovtsev:2005dd}, anti-de Sitter QCD~%
\cite{Katz:2005ir}, etc.

In the present paper we study the decays of the $2^{++}$ tensor nonet into
two pseudoscalars $T \to PP$ and into two photons $T \to \gamma\gamma$
within a chiral approach evaluated at the tree-level. We extend the analysis
to the kinematically allowed strong decay modes into a pseudoscalar and a
vector meson $T\rightarrow PV$ and to the corresponding radiative decay
modes $T\rightarrow P\gamma$.

The basic chiral Lagrangian for the tensor mesons is presented in Refs.~\cite%
{Bellucci:1994eb}-\cite{Chow:1997sg}. In particular, in Refs.~\cite%
{Bellucci:1994eb}-\cite{Donoghue:1988ed} the contribution of the tensor
meson resonances to the low energy coupling constants is evaluated following
the idea suggested in the case of pseudoscalar, scalar, vector and axial
resonances~\cite{Ecker:yg}. In Ref.~\cite{Chow:1997sg} the attention is
focused on the mass spectrum of the tensor mesons. Various decay properties
are analyzed in the framework of tensor-meson dominance (TMD)~ in
combination with vector-meson dominance (VDM)~\cite%
{Renner:1971sj,Suzuki:1993zs,Oh:2003aw}. A study on the phenomenological
properties of $f_{2}(1270)$ and $f_{2}^{\prime }(1525)$ is performed in~\cite%
{Li:2000zb}.

Here we intend to study the decay properties of the full tensor meson nonet
by performing a fit of the free parameters in the chiral Lagrangian to the
available data~\cite{Eidelman:2004wy}. We then turn our attention to some
properties of the tensor glueball with an expected mass of about $\sim $ $%
2.2 $ GeV~\cite{Morningstar:2003ew}. We evaluate the two-pseudoscalar and
two-vector decay ratios as following from the simplest form of the
interaction Lagrangian and discuss some possible candidates.

In Refs.~\cite{Giacosa:2005qr,Giacosa:2005zt} the strong and radiative
decays of the scalar quarkonia nonet supplemented by an intruding glueball
state have been evaluated in a chiral approach. The main difficulty of the
previous study is a rigorous justification of the chiral approach in the
mass region between $1$ and $2$ GeV. At the same time the experimental
situation concerning the scalar sector is not yet complete~\cite%
{Eidelman:2004wy}. The tensor meson sector offers a possibility to test a
tree-level calculation for p-wave states within a chiral approach in the
energy region above $1$ GeV. The $2^{++}$ tensor glueball is expected to
have a mass $\sim 2.2$ GeV~\cite{Morningstar:2003ew} as predicted by lattice
calculation. Also, no significant mixing with the ground state quarkonia is
expected, as the phenomenological study of~\cite{Carvalho:2000ir} confirms.
The quarkonia-mixing between the nonstrange $\overline{n}n$ and the strange $%
\overline{s}s$ component is small, generating an almost ideally mixed nonet.
Flavor mixing driven by instantons, which is large in the pseudoscalar
sector, unknown but possibly large in the scalar one~\cite{Giacosa:2005zt}),
does not affect the tensor mesons.

The interest in hadronic resonance physics at an energy scale larger than $1$
GeV is growing and the attempts to describe mesonic states in chiral
approaches become more numerous. For instance, in~\cite{Roca:2003uk} the
radiative $PV$ decays of axial states (with mass around $1.3$ GeV) were
evaluated. At higher energy scales the calculation of higher order
corrections or possible final state interaction poses a problem to a chiral
approach. However, the chiral Lagrangian with tree-level evaluations
presents a useful phenomenological tool for the description of high-lying
resonances.

The paper is organized as follows. In the next section we describe the
chiral Lagrangian for the $PP$ and $\gamma \gamma $ decays of tensor mesons.
In Sec.3 we perform a fit to the known experimental widths and we discuss
the $\gamma\gamma$, $PV$ and $P\gamma$ transitions. In Sec.4 we consider the
decays of the unmixed tensor glueball with a mass around 2.2~GeV and discuss
some physical resonances. Finally, in Sec.5 we give our conclusions.

\section{Effective Lagrangian for decays of tensor mesons}

The effective Lagrangian describing the strong and electromagnetic decays of
tensor mesons $f_2(1270)$, $f_2^\prime(1525)$, $a_2(1320)$ and $%
K_2^\ast(1430)$ is given by (see detailed discussion in Refs.~\cite%
{Bellucci:1994eb,Toublan:1995bk,Donoghue:1988ed,Chow:1997sg,Suzuki:1993zs}) 
\begin{eqnarray}
\mathcal{L}_{\mathrm{eff}}^{T} &=&\frac{F^{2}}{4}\left\langle D_{\mu
}U\,D^{\mu}U^{\dagger }+\chi _{+}\right\rangle - \frac{1}{4} \langle 
\mathcal{T}_{\mu\nu}D^{\mu \nu ;\rho \sigma} \mathcal{T}_{\rho \sigma }
\rangle - \frac{1}{4} \langle {\mathcal{V}}_{\mu\nu} {\mathcal{V}}^{\mu\nu}
- 2 M_V^2 {\mathcal{V}}_\mu {\mathcal{V}}^\mu \rangle + \,\,\mathcal{L}_{%
\mathrm{mix}}^{P} +\,\mathcal{L}_{\mathrm{mix}}^{T}  \nonumber \\
&+&c_{TPP}^{8}\,\left\langle \mathcal{T}_{\mu \nu }^{octet}\,
\Theta_{P}^{\mu \nu }\right\rangle +\frac{c_{TPP}^{0}}{\sqrt{3}}\, T_{\mu
\nu}^{0}\left\langle \Theta _{P}^{\mu \nu }\,\right\rangle + c_{T\gamma
\gamma }\,\left\langle \, \mathcal{T}_{\mu\nu }\Theta_{\gamma }^{\mu \nu
}\right\rangle  \nonumber \\
&+&ic_{TPV}\left\langle \mathcal{T}^{[\mu \nu ] \alpha }[\widetilde{\mathcal{%
V}}_{\mu \nu },\partial _{\alpha } \mathcal{P}]\right\rangle \, + \,
ic_{TP\gamma}\left\langle \mathcal{T}^{[\mu \nu] \alpha }[Q\widetilde{F}%
_{\mu \nu },\partial _{\alpha }\mathcal{P}] \right\rangle \,,  \label{L_eff}
\end{eqnarray}
where the nonets of tensor, vector and pseudoscalar mesons are defined as 
\begin{equation}
\mathcal{T}_{\mu \nu }=\frac{1}{\sqrt{2}}\sum_{i=0}^{8} T_{\mu
\nu}^{i}\lambda _{i}\,=\mathcal{T}_{\mu \nu }^{octet} +T_{\mu \nu }^{0}\frac{%
\lambda_{0}}{\sqrt{2}}\,, \quad \text{ }\mathcal{V}_\mu=\frac{1}{\sqrt{2}}
\sum_{i=0}^{8}V^{i}_\mu\lambda_{i}\,, \quad \text{ }\mathcal{P}=\frac{1}{%
\sqrt{2}}\sum_{i=0}^{8}P^{i}\lambda_{i}\,.  \label{tpnonet}
\end{equation}
Here and as follows the symbols $\left\langle \ldots \right\rangle$\,, $[
\ldots ]$ and $\{ \ldots \}$ denote the trace over flavor matrices, the
commutator and anticommutator, respectively.

The constants $c_{TPP}^{8},$ $c_{TPP}^{0},$ $c_{T\gamma\gamma },$ $c_{TPV}$
and $c_{TP\gamma }$ define the coupling of tensor fields to photons,
pseudoscalar and vector mesons. We indicate the strong decays of the octet
(coupling $c_{TPP}^{8}$) and the singlet (coupling $c_{TPP}^{0}$) states
separately. However, we do not expect a large violation of the condition $%
c_{TPP}^{8} =c_{TPP}^{0}$ predicted in the large $N_c$ limit.

We use the standard notations for the basic blocks of the ChPT Lagrangian~%
\cite{Weinberg:1978kz,Gasser:1983yg}: $U=u^{2}=\exp (i\mathcal{P}\sqrt{2}/F)$
is the chiral field collecting pseudoscalar fields in the exponential
parametrization, $D_{\mu}$ denotes the chiral and gauge-invariant
derivative, $u_{\mu }=iu^{\dagger}D_{\mu }Uu^{\dagger }$ is the chiral
field, \hspace*{0.1cm}$\chi_{\pm}=u^{\dagger}\chi u^{\dagger}\pm u
\chi^{\dagger}u\,, \hspace*{0.1cm}\chi =2B(s+ip),\,\,\, s=\mathcal{M}+\ldots
\,,$ $\mathcal{M}=\mathrm{diag}\{\hat{m},\hat{m},m_{s}\}$ is the mass matrix
of current quarks (we restrict to the isospin symmetry limit with $%
m_{u}=m_{d}=\hat{m}$)\,, $B$ is the quark vacuum condensate parameter and $F$
is the pseudoscalar meson decay constant.

The term $D^{\mu \nu ;\rho \sigma}$ is the inverse propagator of the tensor
fields and is given by~\cite{Bellucci:1994eb}: 
\begin{eqnarray}
D^{\mu \nu ;\rho \sigma } &=&\left( \square +M_{\mathcal{T}}^{2}\right) %
\left[ \frac{1}{2}\left( g^{\mu \rho }g^{\nu \sigma }+g^{\nu \rho }
g^{\mu\sigma }\right) -g^{\mu \nu }g^{\rho \sigma }\right] +g^{\rho
\sigma}\partial ^{\mu }\partial ^{\nu }+g^{\mu \nu } \partial ^{\rho
}\partial^{\sigma }  \nonumber \\
&-&\frac{1}{2}\left( g^{\nu\sigma }\partial ^{\mu }\partial ^{\rho } +g^{\nu
\rho }\partial ^{\mu}\partial ^{\sigma } +g^{\mu \sigma }\partial ^{\nu
}\partial ^{\rho }+g^{\mu \rho }\partial ^{\nu }\partial ^{\sigma }\right)
\,,  \label{dmunurhosig}
\end{eqnarray}
where $M_{\mathcal{T}}$ is the tensor nonet mass.

The tensors $\Theta _{P}^{\mu \nu }$ and $\Theta _{\gamma }^{\mu \nu}$ are
constructed with the use of chiral and electromagnetic fields: 
\begin{eqnarray}
\Theta _{P}^{\mu \nu } &=&\frac{F^2}{4} \, \{ u^{\mu }\, , \,u^{\nu}\} - \, 
\frac{F^2}{2} \, g^{\mu \nu } \left(u^{\sigma }u_{\sigma }+\chi _{+}\right) ,
\nonumber \\
\Theta _{\gamma }^{\mu \nu } &=&F_{\alpha }^{+\mu }\,F^{+\alpha \nu }+ \frac{%
1}{4}g^{\mu \nu }F^{+\rho \sigma }F_{\rho \sigma }^{+},
\end{eqnarray}
where $F_{\mu \nu }^{+}\,=\,u^{\dagger }F_{\mu \nu }Qu+uF_{\mu \nu}
Qu^{\dagger }$\,, \,\, $F_{\mu \nu } = \partial_\mu A_\nu - \partial_\nu
A_\mu$ is the stress tensor of the electromagnetic field and $Q=e\,\mathrm{%
diag}\{2/3,-1/3,-1/3\}$ is the quark charge matrix.

The tensors $\mathcal{V}^{\mu\nu}$, $\mathcal{T}^{[\mu\nu]\alpha}$ and the
dual tensors $\widetilde{\mathcal{V}}_{\mu\nu}$, $\widetilde{F}_{\mu\nu}$
are defined as 
\begin{eqnarray}
\mathcal{V}^{\mu\nu} &=&\partial^\mu \mathcal{V}^{\nu} - \partial^\nu%
\mathcal{V}^{\mu}\,, \quad \mathcal{T}^{[\mu\nu]\alpha} \,=\,\partial^{\mu}%
\mathcal{T}^{\nu \alpha} -\partial^{\nu }\mathcal{T}^{\mu \alpha }\,, 
\nonumber \\
\widetilde{\mathcal{V}}_{\mu \nu } &=& \frac{1}{2}\varepsilon _{\mu \nu
\rho\sigma } \mathcal{V}^{\rho \sigma}\,, \quad\quad\quad \, \widetilde{F}%
_{\mu \nu } \, = \, \frac{1}{2}\varepsilon_{\mu \nu \rho\sigma} F^{\rho
\sigma} \,.
\end{eqnarray}

We refer to~\cite{Giacosa:2005zt,Ecker:1988te,Venugopal:1998fq} for the
discussion of the term $\mathcal{L}_{\mathrm{mix}}^{P}$ in the Lagrangian,
which contains the pseudoscalar masses and the pseudoscalar mixing. As a
result, the physical states are expressed in terms of the pseudoscalar octet
and singlet states $P^{0}$ and $P^{8}$: 
\begin{eqnarray}
\eta \,=\,P^{8}\cos \theta_{P}\,-\,P^{0}\sin\theta_{P}\,, \hspace*{0.25cm}
\eta^{\prime }\,=P^{8}\,\sin \theta_{P}\,\ +\,P^{0}\,\cos\theta_{P}\,,
\end{eqnarray}
where the pseudoscalar mixing angle reads $\theta _{P}=-9.95^{\circ }$ at
tree-level~\cite{Giacosa:2005zt}.

Here we restrict to the tree-level evaluation, we therefore consistently use
the corresponding tree-level result of $\theta _{P}=-9.95^{\circ }$. In the
present approach we do not include the neutral pion when considering mixing
in the pseudoscalar sector, because we work in the isospin limit. This
mixing is small and can be safely neglected when studying the decay of
tensor resonances into two pseudoscalars. Similarly, for all pseudoscalar
mesons we use the unified leptonic decay constant $F$, which is identified
with the pion decay constant $F=F_{\pi }=92.4$~MeV. A more accurate analysis
including higher orders should use the individual couplings of the
pseudoscalar mesons (for a detailed discussion see Refs.~\cite%
{Weinberg:1978kz,Gasser:1983yg}).

The splitting of the nonet masses and the singlet-octet mixing are generated
by the Lagrangian $\,\mathcal{L}_{\mathrm{mix}}^{T}$ (see~\cite%
{Chow:1997sg,Cirigliano:2003yq}). As a result the physical isoscalar tensor
states $f_{2}\equiv f_{2}(1270)$ and $f_{2}^{\prime }\equiv f_{2}^{\prime
}(1525)$ are expressed in terms of the octet $T^{8}$ and singlet $T^{0}$
components by the tensor mixing angle $\theta_{T}$ (covariant indices $\mu
\nu $ are suppressed): 
\begin{equation}
f\,_{2}=\,T^{0}\,\cos \theta _{T}\,+\,T^{8}\sin \theta _{T}\,, \hspace*{%
0.25cm} f_{2}^{\prime }\,=-\,T^{0}\,\sin \theta _{T}\,\ +\,T^{8}\,
\cos\theta_{T}\,.  \label{thetat}
\end{equation}
The expressions for the two-pseudoscalar decays are derived from the
Lagrangian of Eq.~(\ref{L_eff}) and are listed in the Appendix~A. Note, that
these decays have been discussed previously in Ref.~\cite{Lee:1978py}. We
also report the analytical expressions for the decay rates of the isovector $%
a_2$ and isodoublet $K_2^\ast$ states in the Appendix~A.

The physical vector mesons $\omega $ and $\phi $ are given by (covariant
index $\mu $ understood):%
\begin{equation}
\omega =\,V^{0}\,\cos \theta _{V}\,+\,V^{8}\sin \theta _{V}\,, \hspace*{%
0.25cm} \phi \,=-\,V^{0}\,\sin \theta _{V}\,\ +\,V^{8}\,\cos \theta _{V}\,.
\end{equation}
The vector meson mixing angle $\theta _{V}$ is found to be $39^{\circ }$ in~%
\cite{Cirigliano:2003yq} (the value we use), not far from the ideal mixing
angle of $\theta _{V_{I}}=35.3^{\circ };$ this in turn means that $\omega
\sim \sqrt{1/2}(\overline{u}u+\overline{d}d) \equiv \overline{n}n$ and $\phi
\,\sim \overline{s}s\,.$ Here we describe vector mesons in terms of vector
fields. Alternatively, vector mesons can be represented in terms of
antisymmetric tensor fields which is most convenient for constructing chiral
Largrangians involving vector mesons and their couplings to pseudoscalar
mesons, baryons and photons~\cite%
{Gasser:1983yg,Ecker:1988te,Borasoy:1995ds,Cirigliano:2003yq}.

\section{Results}

\subsection{Two-pseudoscalar decays}

\label{twopseudoscalars}

In this section we perform a fit to the two-body pseudoscalar $PP$ decays of
tensor mesons, in particular the $\pi \pi ,$ $\overline{K}K$, $\eta \eta $
modes for $f_{2}\equiv f_{2}(1270)$ and $f_{2}^{\prime}\equiv f_{2}^{\prime
}(1525),$ the $\overline{K}K,$ $\eta \pi ,$ $\eta ^{\prime }\pi $ modes for $%
a_{2}(1370)$ and the $\overline{K}K$ mode for $K_{2}^{\ast }(1430)$ (the $%
K\eta $ mode is not considered in the fit, see below). The corresponding
experimental results, as deduced from~\cite{Eidelman:2004wy}, are reported
in Table 1. In case of a asymmetric error the largest value is used. The
measured partial decay width $\Gamma _{K_{2}^{\ast }\rightarrow K\eta }/
\left(\Gamma _{K_{2}^{\ast}}\right) _{tot}=1.5_{-1.0}^{+3.4}\times 10^{-3}$
does not allow this procedure, we therefore do not include it in the fit
directly but compare later.

For the partial widths of the various states listed in Table 1 we use the
full widths reported in~\cite{Eidelman:2004wy}. For $K_{2}^{\ast }(1430)$ an
average over the neutral and the charged $K_{2}^{\ast }(1430)$ widths is
performed, finding: $\left(\Gamma_{K_{2}^{\ast}}\right)_{tot}=103.75\pm 3.85$
MeV.

The free parameters entering in the expressions for the $T\rightarrow PP$
decays are the two decay strengths $c_{TPP}^{8}$ and $c_{TPP}^{0}$
introduced in Eq.~(\ref{L_eff}) and the tensor mixing angle $\theta_{T}$
(see Eq.~(\ref{thetat})). The tensor meson masses are taken from Ref.~\cite%
{Eidelman:2004wy}: $M_{f_{2}}=1275.4\pm 1.2$ MeV, $M_{a_{2}}=1318.3\pm 0.6$%
~MeV, $M_{K_{2}^{\ast}}=1429\pm 1.4$~MeV (average over the neutral and the
charged states) and $M_{f_{2}^{\prime }}=1525\pm $ $5$~MeV.

A $\chi^2$ minimum is obtained for the following values: 
\begin{eqnarray}  \label{fitparam}
c_{TPP}^{8}=0.0353 \text{ GeV} \,, \; c_{TPP}^{0}=0.0410 \text{ GeV} \,, \;
\theta_{T}=28.78^{\circ} \;\; \mathrm{with} \; \; \chi _{tot}^{2}=18.496 \,.
\end{eqnarray}
The singlet-octet ratio $y_{TPP} = c_{TPP}^0/c_{TPP}^8 = 1.161$ is close to
unity, as expected from the strong coupling limit. The tensor mixing angle $%
\theta_{T}=28.78^{\circ}$ is not far from the ideal mixing angle $%
\theta_{T_{I}}=35.3^{\circ }$. In the chiral study of \cite%
{Cirigliano:2003yq} the value $\theta_{T}=32^{\circ }$ is obtained, while in
the phenomenological study of~\cite{Li:2000zb} the slightly smaller tensor
angle $\theta_{T}=28.17^{\circ }$ is found.

The mixing matrices connecting the physical states $f_2,$ $f_2^\prime$ to
the bare ones, $T^{0}$ and $T^{8},$ or to $\overline{n}n\equiv 1/\sqrt{2}(%
\overline{u}u+\overline{d}d)$ and $\overline{s}s$ read explicitly: 
\begin{equation}
\left( 
\begin{array}{c}
f_{2} \\ 
f_{2}^{\prime }%
\end{array}
\right) =\left( 
\begin{array}{cc}
0.876 & 0.481 \\ 
-0.481 & 0.876%
\end{array}
\right) \left( 
\begin{array}{c}
T^{0} \\ 
T^{8}%
\end{array}
\right) =\left( 
\begin{array}{cc}
0.993 & 0.113 \\ 
0.113 & -0.993%
\end{array}
\right) \left( 
\begin{array}{c}
\overline{n}n \\ 
\overline{s}s%
\end{array}
\right) .  \label{mixmatr}
\end{equation}
The physical states are very close to pure $\overline{n}n$ and $\overline{s}%
s $ configurations, which is particularly evident from the small $f_2^\prime
\to \pi\pi$ partial decay widths. Contrary to the pseudoscalar sector and
perhaps to the scalar one~\cite{Giacosa:2005zt,Minkowski:2002nf} a large
flavor mixing in the tensor nonet is not expected.

The fit results are summarized in Table 1.

\begin{center}
\textbf{Table 1.} Decay properties of tensor mesons.

\vspace*{0.5cm} 
\begin{tabular}{|l|l|l|l|}
\hline
Mode & Experiment (MeV) & Theory (MeV) & $\chi _{i}^{2}$ \\ \hline
$\Gamma _{f_{2}\rightarrow \pi \pi }$ \thinspace & \hspace*{0.6cm}$157.0\pm
7.6$ & \hspace*{0.6cm}$153.51$ & $0.210$ \\ \hline
$\Gamma _{f_{2}\rightarrow \bar{K}K}$ \thinspace \thinspace & \hspace*{0.6cm}
$8.5\pm 0.9$ & \hspace*{0.6cm} $9.15$ & $0.526$ \\ \hline
$\Gamma _{f_{2}\rightarrow \eta \eta }$ \thinspace & \hspace*{0.6cm}$0.83\pm
0.20$ & \hspace*{0.6cm} $0.80$ & $0.023$ \\ \hline
$\Gamma _{f_{2}^{\prime }\rightarrow \pi \pi }$ \thinspace \thinspace & 
\hspace*{0.6cm} $0.60\pm 0.16$ & \hspace*{0.6cm} $0.55$ & $0.102$ \\ \hline
$\Gamma _{f_{2}^{\prime }\rightarrow \bar{K}K}$ \thinspace \thinspace & 
\hspace*{0.6cm} $64.8\pm 7.6$ & \hspace*{0.6cm} $41.64$ & $9.288$ \\ \hline
$\Gamma _{f_{2}^{\prime }\rightarrow \eta \eta }$ & \hspace*{0.6cm} $7.5\pm
2.9$ & \hspace*{0.6cm} $6.49$ & $0.121$ \\ \hline
$\Gamma _{a_{2}\rightarrow \bar{K}K}$ \thinspace \thinspace & \hspace*{0.6cm}
$5.2\pm 1.1$ & \hspace*{0.6cm} $6.64$ & $1.716$ \\ \hline
$\Gamma _{a_{2}\rightarrow \eta \pi }$ & \hspace*{0.6cm} $15.5\pm 2.0$ & 
\hspace*{0.6cm} $18.42$ & $2.134$ \\ \hline
$\Gamma _{a_{2}\rightarrow \eta ^{\prime }\pi }$ & \hspace*{0.6cm} $0.57\pm
0.12$ & \hspace*{0.6cm} $0.80$ & $3.652$ \\ \hline
$\Gamma _{K_{2}^{\ast }\rightarrow \bar{K}K}$ & \hspace*{0.6cm} $51.8\pm 3.2$
& \hspace*{0.6cm} $49.08$ & $0.722$ \\ \hline
$\chi _{tot}^{2}$ & \hspace*{1.4cm} - & \hspace*{0.95cm} - & $18.496$ \\ 
\hline
\end{tabular}
\end{center}

The description of the experimental data ($\chi _{tot}^{2}/N=1.85$) is good.
The largest contribution to $\chi ^{2}$ comes from an underestimate of the $%
\overline{K}K$ mode for the $f_{2}^{\prime }(1525)$ resonance.

The theoretical prediction for the branching ratio $K_{2}^{\ast }\rightarrow
K\eta $ is 
\begin{eqnarray}
\Gamma _{K_{2}^{\ast }\rightarrow K\eta }/ \left( \Gamma
_{K_{2}^{\ast}}\right) _{tot}=3.93\,\times 10^{-3}
\end{eqnarray}
which within errors is in agreement with the corresponding experimental
value of $1.5_{-1.0}^{+3.4}\times 10^{-3}$.

\subsection{Two-photon decays}

\label{twophoton}

We now turn to $\gamma \gamma $ decays of the isoscalar and neutral
isovector tensor states. The ratios $\Gamma _{f_{2}^{\prime }\rightarrow
\gamma \gamma }/\Gamma _{f_{2}\rightarrow \gamma \gamma }$ and $%
\Gamma_{a_{2}\rightarrow \gamma \gamma }/ \Gamma _{f_{2}\rightarrow \gamma
\gamma }$ do not depend on the strength $c_{T\gamma \gamma }$ in (\ref{L_eff}%
) and numerically read 
\begin{equation}
\Gamma _{f_{2}^{\prime }\rightarrow \gamma \gamma }/
\Gamma_{f_{2}\rightarrow \gamma \gamma }=0.046\,, \quad \Gamma
_{a_{2}\rightarrow \gamma \gamma }/\Gamma _{f_{2}\rightarrow \gamma \gamma
}=0.378,
\end{equation}
where the tensor mixing angle $\theta_{T}=28.17^{\circ }$ is used as fixed
by the fit (see~(\ref{fitparam})). The corresponding experimental values are~%
\cite{Eidelman:2004wy}: 
\begin{eqnarray}
\left( \Gamma _{f_{2}^{\prime }\rightarrow \gamma \gamma }/
\Gamma_{f_{2}\rightarrow \gamma \gamma }\right) _{\exp } &=&0.031\pm 0.010,%
\text{ }  \nonumber \\
\left( \Gamma _{a_{2}\rightarrow \gamma \gamma }/\Gamma _{f_{2}
\rightarrow\gamma \gamma }\right) _{\exp } &=&0.383\pm 0.057\,.
\end{eqnarray}
Again, the small value of $\Gamma_{f_{2}^{\prime}\rightarrow \gamma \gamma
}/\Gamma _{f_{2}\rightarrow \gamma \gamma }$ is extremely sensitive to the
precise value of the tensor mixing angle.

In~\cite{Suzuki:1993zs,Oh:2003aw,Cotanch:2005ja} a method is used to fix the
strength of the two-photon decays: the tensor meson dominance (TMD)
hypothesis allows to determine the coupling of tensor mesons to vector
mesons and the subsequent application of vector meson dominance (VMD) allows
to deduce the two-photon decay rates. In the present work we do not intend
to perform a systematic study of the $TVV$ coupling (and therefore of VMD).
The decay into two vectors is generally not kinematically allowed for a
ground state tensor meson. The presence of a $4\pi $ decay mode for the
state $f_{2}$ is indeed an indication of a contribution of virtual vector
mesons, which then decay into pions. The calculation of such contributions
is possible by taking properly into account the finite widths of the
resonances and the corresponding virtual states, but, although being an
interesting subject, will not be analyzed in the present work. As indicated
in~\cite{Suzuki:1993zs}, the results obtained by applying VMD compare only
moderately to the data.

\subsection{Vector-Pseudoscalar decays}

\label{tvp}

The isovector state $a_{2}$ decays into $\rho \pi $ ($\overline{K}%
K^{\ast}(892)$ is not kinematically allowed), the isodoublets $K_{2}^{\ast }$
into $\overline{K}K^{\ast }(892),$ $K\rho $ and $K\omega .$ By fixing the
strength parameter $c_{TPV}$ to reproduce the decay rate of $%
a_{2}\rightarrow \rho\pi $ we can predict the other three rates (see Table
2). The presented lowest-order results for the decay of $K_{2}^{\ast}(1430)$%
, depending only on one free parameter and on flavor symmetry, are rather
good.

\begin{center}
\textbf{Table 2.} $T\rightarrow PV$ decays

\vspace*{0.5cm} 
\begin{tabular}{|l|l|l|}
\hline
Quantity & Experiment (MeV) & Theory (MeV) \\ \hline
$\Gamma _{a_{2}\rightarrow \pi \rho }$ & $75.0\pm 6.4$ & $75.0$ (fixed) \\ 
$\Gamma _{K_{2}^{\ast }\rightarrow \pi K^{\ast }(892)}$ & $24.5\pm 1.4$ & $%
28.97$ \\ 
$\Gamma _{K_{2}^{\ast }\rightarrow K\rho }$ & $8.6\pm 1.0$ & $7.40$ \\ 
$\Gamma _{K_{2}^{\ast }\rightarrow K\omega }$ & $2.86\pm 0.87$ & $2.64$ \\%
[1mm] \hline
\end{tabular}
\end{center}

\subsection{Pseudoscalar-photon decays}

The decay rates $a_{2}^{\pm }\rightarrow \pi ^{\pm }\gamma $ and $%
K_{2}^{\ast \pm }\rightarrow K^{\pm }\gamma $ depend on the coupling
constant $c_{TP\gamma },$ but their ratio does not and reads: 
\begin{equation}
\frac{\Gamma _{K_{2}^{\ast \pm }\rightarrow K^{\pm }\gamma }} {%
\Gamma_{a_{2}^{\pm }\rightarrow \pi ^{\pm }\gamma }}=0.83\,,
\end{equation}
which we compare to the experimental value of 
\begin{equation}
\left( \frac{\Gamma _{K_{2}^{\ast \pm }\rightarrow K^{\pm }\gamma }}{\Gamma
_{a_{2}^{\pm }\rightarrow \pi ^{\pm }\gamma }}\right) _{\exp }=\frac{%
0.236\pm 0.056}{0.287\pm 0.047} = 0.82\pm 0.29\,,
\end{equation}
which is in good agreement.

As already indicated in the previous subsection, we do not intend to
evaluate the radiative decays via vector-meson-dominance. Note, however,
that the final states $PV$ are in a $d$-wave, thus implying a fifth power of
the relative momentum (see Appendix A) and therefore the kinematical
contribution dominates. A naive application of VMD leads therefore to an
overestimate of the $P\gamma $-width.

\section{Decays of tensor glueball}

\subsection{$PP$ and $VV$ decay ratios}

\label{ppvvratios}

According to Lattice QCD the lightest glueball has quantum numbers $%
J^{PC}~=~0^{++}$ and a mass of about $1.5$ GeV~\cite{Morningstar:2003ew},
which likely mixes with the nearby quarkonia states generating the three
scalar-isoscalar resonances $f_{0}(1370),$ $f_{0}(1500)$ and $f_{0}(1710).$
This mixing scenario, although not unique (see~\cite%
{Amsler:2004ps,Minkowski:2002nf} and Refs. therein), has been analyzed in
various ways: at a composite level in the quantum mechanical studies of~\cite%
{Amsler:1995td,Strohmeier-Presicek:1999yv,Close:2001ga}, in the Lattice
study of~\cite{Lee:1999kv} and by means of a chiral approach~\cite%
{Giacosa:2005qr,Giacosa:2005zt}. Also, attempts at a microscopic quark-gluon
level as in~\cite{Giacosa:2004ug} (and Refs. therein) have been performed.

Lattice QCD sets the tensor glueball mass around $2.2$ GeV~\cite%
{Morningstar:2003ew}. A significant mixing with the tensor ground state
mesons analyzed in the previous section can be excluded due to the large
mass difference (see the study of~\cite{Carvalho:2000ir}). On the other
hand, a mixing with excited isoscalar-tensor quarkonia states lying in the
same energy region is possible, but unfortunately very difficult to deduce:
the experimental informations are scarce and the application of theoretical
methods is only partially reliable.

In the following we evaluate the two-pseudoscalar and two-vector decay
ratios for a hypothetical flavour-blind composite state with a mass of about 
$\sim$ 2.2 GeV, where glueball-quarkonia mixing is neglected: although one
should be aware of such an eventuality, an analysis of mixing requires a
certain amount of data to deduce the wave-function contributions. Indeed,
the possibility of a small glueball-quarkonia mixing in the tensor sector
can be conceived, as we discuss in the next subsection.

Neglecting phase space and flavor symmetry breaking, the two-pseudoscalar
ratios for a flavor-blind tensor glueball $G_{\mu \nu }$ follow the
well-known pattern: 
\begin{equation}  \label{GPP_ratios}
\pi \pi :\overline{K}K:\eta \eta :\eta \eta ^{\prime }: \eta^{\prime
}\eta^{\prime }=3:4:1:0:1\,.
\end{equation}
By introducing the glueball as a flavor-blind composite field we write down
the effective Lagrangian describing its decays into two photons and two
vector mesons: 
\begin{eqnarray}  \label{Leff_G}
\mathcal{L}_{\mathrm{eff}}^{G} \, = \, c_{GPP} \, G_{\mu\nu} \, \left\langle
\, \Theta_{P}^{\mu \nu }\,\right\rangle \, + \, c_{GVV} \, G_{\mu \nu } \,
\left\langle \, \mathcal{V}^{\mu } \, \mathcal{V}^{\nu } \, \right\rangle \,.
\end{eqnarray}
Fixing the tensor glueball mass as $M_{G_{2}}=2.2$ GeV (in accord with
Lattice) the $PP$-ratios become: 
\begin{equation}
\pi \pi :\overline{K}K:\eta \eta :\eta \eta^{\prime }: \eta^{\prime}
\eta^{\prime }=1:0.79:0.17:0:0.001\,.  \label{ppratios}
\end{equation}
Compared to the flavor ratios of (\ref{GPP_ratios}), here the $\pi\pi$ is
enhanced and $\eta^{\prime }\eta ^{\prime }$ is highly suppressed because of
available phase space. The $\eta \eta ^{\prime }$ is still zero, since no
flavor-breaking term is present in~(\ref{Leff_G}).

Similarly, for the two-vector decay ratios one has due to flavor symmetry
considerations 
\begin{equation}
\rho \rho :\overline{K}^{\ast }K^{\ast }:\omega \omega :\omega \phi :\phi
\phi =3:4:1:0:1\,.  \label{VV-ratios_naive}
\end{equation}%
Inclusion of the phase space correction changes the $VV$-ratios~(\ref%
{VV-ratios_naive}): 
\begin{equation}
\rho \rho :\overline{K}^{\ast }K^{\ast }:\omega \omega :\omega \phi :\phi
\phi =1:0.84:0.32:0:0.11\,.  \label{vvratios}
\end{equation}%
Note, in section~\ref{twopseudoscalars} we did not consider the decays of
the ground-state tensor mesons into pair of vector mesons, because such a
decay is generally not kinematically allowed. This kind of approach has been
performed in~\cite{Suzuki:1993zs,Oh:2003aw}, where TMD is employed to deduce
the $TVV$ interaction. This approach is the starting point of the analysis
of~\cite{Cotanch:2005ja}, where the two-vector and the radiative decays of a
gluonic state with a mass $M\gtrsim $ $2$ GeV are studied. Their results are
similar to (\ref{vvratios}), apart from the $\omega \phi $ mode, large in 
\cite{Cotanch:2005ja} and zero in our approach : the $\omega \phi $ mode is
zero (independent on the choice of the vector mixing angle $\theta _{V}$),
because a flavor singlet cannot decay into a singlet and an octet (or a
mixture of those) because of $U(3)$ flavor-symmetry. This phenomenon is
completely analogous to the predicted zero decay mode of a scalar (or a
tensor) glueball into $\eta \eta ^{\prime }$. In~\cite{Giacosa:2005zt} a
possible $\eta \eta ^{\prime }$ decay of the (unmixed) scalar glueball is
generated by a $U(3)$ flavor breaking contribution. We also refer to the
work of~\cite{Anisovich:2004vj}, where the flavor coefficients for the
tensor glueball into two vectors are reported (leading-order results in the
flavor symmetry limit): the $\eta \eta ^{\prime }$ and the $\omega \phi $
modes are forbidden because of $U(3)$ flavor-symmetry. A breaking of this
symmetry is possible, but then also the other modes are affected~\cite%
{Giacosa:2005zt,Anisovich:2004vj}. For completeness we consider such a
possibility of the $U(3)\rightarrow SU(3)$ breaking when the nonet of vector
mesons is splitted into octet and singlet states. Then we have different
couplings of the tensor glueball to octet ($c_{GVV}^{8}$) and singlet ($%
c_{GVV}^{0}$) states: 
\begin{eqnarray}\label{U3_breaking}
\mathcal{L}_{\mathrm{eff}}^{GVV}\,=\,c_{GVV}^{8}\,G_{\mu \nu }\,\langle 
\mathcal{V}^{\mu \,octet}\,\mathcal{V}^{\nu \,octet}\rangle
\,+\,c_{GVV}^{0}\,G_{\mu \nu }\,V^{\mu \,0}\,V^{\nu \,0}\,.
\end{eqnarray} 
The decay amplitudes are reported in the Table 8 in the Appendix: the $%
\omega \phi $ mode is now allowed and the corresponding amplitude is
proportional to $\left[ 1-c_{GVV}^{0}/\,c_{GVV}^{8}\right] ,$ therefore is a
higher order correction to the large $N_{c}$ limit with $c_{GVV}^{0}=%
\,c_{GVV}^{8}.$ No large violation from this limit is expected, as the
two-pseudoscalar decay modes confirmed (resulting in $%
y_{TPP}=c_{TPP}^{0}/c_{TPP}^{8}=1.161$, see Sec.II). For $%
c_{GVV}^{0}/\,c_{GVV}^{8}\leq 1.56$ we still have a small $\omega \phi $
mode with $\omega \phi /\rho \rho \leq 0.1.$ For $c_{GVV}^{0}/%
\,c_{GVV}^{8}=1.56$ the ratios read 
\begin{eqnarray} 
\rho \rho :\overline{K}^{\ast }K^{\ast }:\omega \omega :\omega \phi :\phi
\phi =1:0.84:0.58:0.1:0.16\,,
\end{eqnarray}
therefore still in qualitative agreement with (\ref{vvratios}).

In order to get $\omega \phi /\rho \rho \approx 1$ as in~\cite%
{Cotanch:2005ja} we have to increase the value of the ratio $%
c_{GVV}^{0}/\,c_{GVV}^{8}$ up to $\approx 2.79,$ implying a large (and
unnatural) difference between the octet and the singlet decay parameters.
For $c_{GVV}^{0}/\,c_{GVV}^{8}=2.79$ we have 
\begin{eqnarray}
\rho \rho :\overline{K}^{\ast }K^{\ast }:\omega \omega :\omega \phi :\phi
\phi =1:0.84:1.41:1.00:0.32\,.
\end{eqnarray}
The $\omega \omega $ mode has also been modified, being very large in this
scenario.

In the end, a strong $\omega \phi $ mode is possible only by introducing a
consistent violation from the large $N_{c}$ limit, corresponding to a
flavor-undemocratic tensor glueball decay. Although such an eventuality
cannot be excluded (see~\cite{Giacosa:2005zt} for the discussion in the
scalar sector, where however the "undemocracy" is a result of a
flavor-symmetry breaking term, affecting also the kaonic decay modes, and
not the octet-singlet splitting) is at the present state of knowledge not
verifiable and in disagreement with other results, where large $N_{c}$ is,
although broken, still approximately valid.

\subsection{Discussion of $f_{J}(2220)$ as a tensor-glueball candidate}

Limiting our study to the mass region $M\gtrsim $ $2$ GeV, the following
isoscalar tensor states are listed in~\cite{Eidelman:2004wy}: $f_{2}(2010),$ 
$f_{2}(2150),$ $f_{J}(2220)$ ($J=0$ or $2$, which still needs to be
settled), $f_{2}(2300)$ and $f_{2}(2340).$

In~\cite{Cotanch:2005ja} the decays of a flavor-blind tensor glueball have
been evaluated for $f_{2}(2010)$ and $f_{2}(2300)$. In~\cite%
{Anisovich:2004vj} the analysis of Regge-trajectories for the tensor states
leads to the interpretation of $f_{2}(2010)$ and the $f_{2}(2300)$ as
dominant $\overline{s}s$ states (note, that a different naming scheme from 
\cite{Eidelman:2004wy} is used in~\cite{Anisovich:2004vj}). According to~%
\cite{Anisovich:2004vj} all the isoscalar tensor states, with the exception
of $f_{J}(2220)$ and a broad tensor state around 2 GeV (not listed in~\cite%
{Eidelman:2004wy}), can be interpreted as quarkonia. The broad state ($%
\Gamma \sim 500$ MeV) around $2$ GeV (mass between $1.7$ and $2.5$ GeV)
found in the analysis of~\cite{Anisovich:2004vj} and denoted as $f_{2}(2000)$%
, but not listed in the compilation of~\cite{Eidelman:2004wy}, is
interpreted as the tensor glueball (see also~\cite{Anisovich:2005kv}). The
debated issue of the full width of the glueball in the scalar sector~\cite%
{Amsler:1995td,Giacosa:2005qr,Giacosa:2005zt,Strohmeier-Presicek:1999yv,Lee:1999kv}
is one of the main questions in the tensor sector as well. Another possible
candidate for the tensor glueball is the very narrow state $f_{J}(2220)$ ($%
\Gamma _{tot}=23_{-7}^{+8}$ MeV), also not lying on the Regge trajectories
explored in~\cite{Anisovich:2004vj}. This possibility is "opposite" to the
broad tensor glueball discussed above (see also the discussion in the end of~%
\cite{Anisovich:2004vj} and Refs. therein). This narrow resonance is in line
with the interpretation of a narrow glueball~\cite{Amsler:2004ps} (we also
refer to the note of Doser in~\cite{Eidelman:2004wy} on~$f_{J}(2220)$).

The branching ratio $\Gamma_{\pi\pi}/\Gamma_{\overline{K}K} = 1.0\pm 0.5$~%
\cite{Eidelman:2004wy} is compatible with~(\ref{ppratios}). The absence of a 
$\gamma\gamma $-signal is also in line with a narrow gluonic state. If the
glueball is broad, also its vector-vector decay modes are expected to be
broad. Therefore in virtue of vector-meson dominance, the two-photon signal
is also expected to be large. The general idea that the two-photon signal
should be small for a glueball is indeed valid only for a narrow glueball.
However, the $\eta \eta^{\prime }$ mode (zero according to the leading-order
results expressed in~(\ref{ppratios})) has been seen, while the $\eta \eta $
mode has not. More precise branching ratios are needed to analyze $%
f_{J}(2220)$ quantitatively, but this resonance has some intriguing
characteristics to be considered as a tensor-glueball candidate.

The properties discussed up to now hold for an unmixed tensor glueball. When
considering $f_{J}(2220)$ as a glueball candidate mixing is therefore
neglected. In the pseudoscalar meson sector the physical states $\eta $ and $%
\eta ^{\prime }$ are close to octet and singlet states, i.e. far from $%
\overline{n}n$ and $\overline{s}s,$ a fact notoriously connected with the $%
U_{A}(1)$-QCD anomaly. A strong mixing among $\overline{n}n$ and $\overline{s%
}s$ is generated, possibly by instantons. A mixing with a pseudoscalar
glueball would then be expected to be large, if the $J^{PC}=0^{-+}$glueball
had a mass in the energy sector below 1 GeV (but lattice places it higher
than $2$ GeV).

The scalar sector is more controversial: a strong glueball-quarkonia mixing
is the starting point of the phenomenological works~\cite%
{Amsler:1995td,Giacosa:2005zt,Close:2001ga,Strohmeier-Presicek:1999yv} 
and has also been verified by 
lattice simulation~\cite{Lee:1999kv,McNeile:2000xx}. 
A strong $\overline{n}n$-$\overline{s}s$ quarkonia mixing (possibly driven
by instantons~\cite{Klempt:2000ud}) has been considered in~\cite%
{Giacosa:2005qr,Giacosa:2005zt}. In the NJL model (\cite{Hatsuda:1994pi} and
Refs. therein) the $U_{A}(1)$ anomaly is introduced by the t' Hooft
interaction term, which affects the pseudoscalar and the scalar mesonic
sectors. A gluonic interaction seems therefore enhanced for the $%
J^{PC}=0^{-+}$ and $J^{PC}=0^{++}$ states, although in the scalar sector
further confirmation is needed. On the other hand, no evidence of enhanced $%
\overline{n}n$-$\overline{s}s$ quarkonia mixing is found in other nonets~%
\cite{Cirigliano:2003yq}; according to~\cite{Anisovich:2004vj}, this fact
holds also for the excited tensor mesons. The physical states are close to
the bare light quark configurations $\overline{n}n$ and $\overline{s}s.$
Following this observation, one can argue that a glueball would not mix
strongly in this case (this point of view is not accepted by~\cite%
{Burakovsky:1999ug}, where the state $f_{J}(2220)$ is studied but is
considered as a broad resonance). It is important to stress that such a
reasoning is qualitative and needs quantitative theoretical analysis, while
at the same time a better experimental understanding of $f_{J}(2220)$ is
required. 

\section{Conclusions}

In this paper we first studied the decay properties of the ground-state
tensor meson nonet: the starting point has been a chiral approach for the
two-pseudoscalar decays, where a fit has been performed to deduce the free
parameters of the employed Lagrangian~(\ref{L_eff}). The tree-level results
are in good agreement with the experimental data. When defining the chiral
tensor-pseudoscalar coupling we splitted the tensor meson octet from the
singlet: the singlet decay strength turns out to be slightly larger than the
octet one. Such a octet-singlet separation is necessary to obtain acceptable
results for the fit because of the precision of the experimental data. As a
result of the fit the tensor mixing angle and the two-photon decay ratios
have been deduced and are in agreement with other approaches and with data. 

We then turned our attention to the pseudoscalar-vector and
pseudoscalar-photon decay modes for tensor mesons, where the simplest
coupling respecting $CPT$-invariance has been considered. The corresponding
theoretical results are in good agreement with data.

As a further step of our study on tensor states we considered the tensor
glueball, described by a flavor-blind composite field with an independent
coupling to two pseudoscalar mesons (the approach is analogous to the
analysis in the scalar sector performed in~\cite%
{Giacosa:2005qr,Giacosa:2005zt})). The two-pseudoscalar ratios are then
presented, where the mass of $2.2$ GeV as obtained on the lattice, has been
used. The two-vector decay ratios are then also analyzed.
The full strength for the tensor glueball decays is unknown and represents
the most interesting and debated issue for glueball decays (not only in the
tensor sector). 
A narrow tensor glueball is discussed in~\cite{Amsler:2004ps}, 
while a broad tensor glueball in~\cite{Anisovich:2004vj}. 

We discussed the narrow resonance $f_{J}(2220)$ ($\Gamma \sim 30$ MeV) as a
possible unmixed tensor-glueball candidate; however, the total spin and the
nature of this state are still controversial~\cite%
{Eidelman:2004wy,Anisovich:2004vj,Burakovsky:1999ug}. The possibility that
glueball-mixing is small in the tensor sector has been discussed, but
further and more quantitative studies are needed in this direction.

\begin{acknowledgments}
This work was supported by the by the DFG under contracts FA67/25-3 and
GRK683. This research is also part of the EU Integrated Infrastructure
Initiative Hadron physics project under contract number RII3-CT-2004-506078
and President grant of Russia "Scientific Schools" No. 1743.2003.
\end{acknowledgments}

\appendix

\section{Expressions for the decay widths}

The generic two-pseudoscalar decay expression of a tensor state reads (see
also Refs.~\cite{Suzuki:1993zs,Bellucci:1994eb,Lee:1978py}): 
\begin{eqnarray}  \label{Decay_TPP}
\Gamma_{TP_{1}P_{2}} \, = \, \alpha_{TP_1P_2} \, \frac{P_{TP_1P_2}^5}{60\,
\pi \, M_{T}^{2}} \, g_{TP_1P_2}^2
\end{eqnarray}
where $P_{TP_1P_2} = \lambda^{1/2}(M_T^2,M_{P_1}^2,M_{P_2}^2)/2 M_T$ is the
three-momentum of the final states (pseudoscalar mesons) in the rest frame
of the decaying initial state (tensor meson) and $\lambda(x,y,z) = x^2 + y^2
+ z^2 - 2 xy - 2 xz - 2 yz$ is the K\"{a}llen triangle function. The $%
\alpha_{TP_{1}P_{2}}$ takes into account the average over spin of the
initial state and the sum over final isospin states with averaging over
initial isospin states (symmetry factors included), $g_{TP_{1}P_{2}}$ is the
effective $TP_{1}P_{2}$ coupling constant which is defined for different
modes as: 
\begin{eqnarray}  \label{L_TPP}
\mathcal{L}_{TP_1P_2} &=& (g_{f_2\pi\pi}
f_2^{\mu\nu}\,+\,g_{f_2^\prime\pi\pi} f_2^{\prime\mu\nu}) \, \partial_\mu%
\vec{\pi} \, \partial_\nu\vec{\pi} \ \, + \, ( \, g_{f_2 \overline K K}
f_2^{\mu\nu} \, + \, g_{f_2^\prime \overline K K} f_2^{\prime\mu\nu} \, ) \,
\partial_\mu K^\dagger \, \partial_\nu K \, \\
[5mm] &+&(g_{f_2\eta\eta} f_2^{\mu\nu} \, + \, g_{f_2^\prime\eta\eta}
f_2^{\prime\mu\nu}) \, \partial_\mu\eta \, \partial_\nu\eta \, + \, \frac{%
g_{a_2 \overline K K}}{\sqrt{2}} \, \partial_\mu K^\dagger \, \vec{a}_2^{\,
\mu\nu} \, \vec\tau \, \partial_\nu K \, + \, \vec{a}_2^{\, \mu\nu} \,
\partial_\mu\vec\pi \, (g_{a_2 \pi \eta} \, \partial_\nu\eta \, + \, g_{a_2
\pi \eta^\prime} \, \partial_\nu\eta^\prime) \,  \nonumber \\
[2mm]  \nonumber \\
&+& \frac{g_{K_2^\ast \pi K}}{\sqrt{2}} \, ( K^{\ast\mu\nu \,\dagger}_{2} \,
\partial_\mu\vec{\pi} \,\, \vec\tau \, \partial_\nu K \, + \, \mathrm{h.c.}
\ ) \, + \, g_{K_2^\ast \eta K} \, ( K^{\ast\mu\nu \,\dagger}_2 \,
\partial_\mu K \, \partial_\nu\eta \, + \, \mathrm{h.c.} \ ) \,  \nonumber
\end{eqnarray}
where $\vec{\pi}$ and $\vec{a}_2$ are the triplets of pions and tensor $a_2$
mesons, $K = (K^+, K^0)$ and $K^\dagger = (K^-, \overline K^0)$ are the
doublets of kaons, $K^{\ast}_2 = (K^{\ast +}_2, K^{\ast 0}_2)$ and $K^{\ast
\dagger}_2 = (K^{\ast -}_2, \overline K^{\ast 0}_2)$ are the doublets of
tensor $K^\ast_2$ mesons.

The results for the parameters involving in Eq.~(\ref{Decay_TPP}) are
reported in Table~3. We introduce the notations: $\tilde g_{TP_1P_2} =
g_{TP_1P_2}/c_{TPP}^8$ is the coupling constant scaled by $c_{TPP}^8$, the
parameter $y=c_{TPP}^0/c_{TPP}^8$ is the ratio of the singlet and octet
couplings, $z_P = (1 + 3 \cos 2\delta_P)/2$, where $\delta_P = \theta_P -
\theta_P^I$ and $\theta_P^I$ is the ideal mixing angle with $\sin\theta_P^I
= 1/\sqrt{3}$.

\vspace*{1cm}

\begin{center}
\textbf{Table 3.} $T\rightarrow PP$ coefficients.

\vspace*{0.5cm}

\begin{tabular}{|l|l|l|l|}
\hline
\hspace*{.6cm}$T$ & $P_{1}P_{2}$ & \hspace*{.2cm}$\alpha_{TP_{1}P_{2}}$ & 
\hspace*{1.7cm}$\tilde g_{TP_1P_2}$ \\ \hline
$f_{2}(1270)$ & $\pi \pi $ & \hspace*{.5cm}$6$ & $\frac{y}{\sqrt{3}} \,
\cos\theta_{T} + \frac{1}{\sqrt{6}} \, \sin\theta_{T}$ \\ \hline
$f_{2}(1270)$ & $\overline{K}K$ & \hspace*{.4cm} $8$ & $\frac{y}{\sqrt{3}}
\, \cos\theta_{T} - \frac{1}{2 \sqrt{6}} \, \sin\theta_{T}$ \\ \hline
$f_{2}(1270)$ & $\eta \eta $ & \hspace*{.4cm} $2$ & $\frac{y}{\sqrt{3}} \,
\cos\theta_{T} - \frac{1}{\sqrt{6}} \, \sin\theta_{T} \, z_P$ \\ \hline
$f_{2}^{\prime }(1525)$ & $\pi \pi $ & \hspace*{.4cm} $6$ & $- \frac{y}{%
\sqrt{3}} \, \sin\theta_{T} + \frac{1}{\sqrt{6}} \, \cos\theta_{T}$ \\ \hline
$f_{2}^{\prime }(1525)$ & $\overline{K}K$ & \hspace*{.4cm} $8$ & $- \frac{y}{%
\sqrt{3}} \, \sin\theta_{T} - \frac{1}{2 \sqrt{6}} \, \cos\theta_{T}$ \\ 
\hline
$f_{2}^{\prime }(1525)$ & $\eta \eta $ & \hspace*{.5cm}$2$ & $- \frac{y}{%
\sqrt{3}} \, \sin\theta_{T} - \frac{1}{\sqrt{6}} \, \cos\theta_{T} \, z_P$
\\ \hline
$a_{2}(1320)$ & $\overline{K}K$ & \hspace*{.4cm} $1$ & \hspace*{2cm}$1$ \\ 
\hline
$a_{2}(1320)$ & $\pi \eta $ & \hspace*{.55cm}$1$ & \hspace*{1cm} $- \sqrt{2}
\sin\delta_{P}$ \\ \hline
$a_{2}(1320)$ & $\pi \eta^\prime $ & \hspace*{.55cm}$1$ & \hspace*{1.45cm}$%
\sqrt{2} \cos\delta_{P}$ \\ \hline
$K_{2}(1430)$ & $\pi K$ & \hspace*{.2cm} $3/2$ & \hspace*{2cm}$1$ \\ \hline
$K_{2}(1430)$ & $\eta K$ & \hspace*{.55cm}$1$ & $- \frac{1}{\sqrt{6}}%
(\cos\theta_{P} + 2 \sqrt{2}\sin\theta_{P})$ \\ \hline
\end{tabular}
\end{center}

\vspace*{.5cm}

The expression for the $PV$ decay width reads~\cite{Lee:1978py}: 
\begin{eqnarray}
\Gamma _{TPV} \, = \, \alpha_{TPV}\, \frac{P_{TPV}^5}{10\pi} \, g_{TPV}^2 \,,
\end{eqnarray}
where $\alpha_{TPV}$ and $\tilde g_{TPV} = g_{TPV}/c_{TPV}$ are given in
Table 4. The effective couplings $g_{TPV}$ are defined as: 
\begin{eqnarray}  \label{L_TVP}
\mathcal{L}_{TVP} &=& \frac{g_{f_2^\prime K K^\ast}}{\sqrt{2}} \, ( \,
f_{2}^{\prime \, [\mu\nu]\alpha} \, i \, \partial_\alpha K^\dagger \, \tilde
K_{\mu\nu}^{\ast} \, \ + \ \mathrm{h.c.} \, )  \nonumber \\[3mm]
&+& \frac{g_{a_2\rho\pi}}{\sqrt{2}}\,\vec{a}_{2}^{\,[\mu\nu]\alpha} \, \cdot
\,[ \partial_\alpha \vec{\pi} \, \times \, \tilde{\vec{\rho}}_{\mu\nu}] \, +
\, \frac{g_{a_2 K K^\ast}}{\sqrt{2}} \, ( \, i \, \partial_\alpha K^\dagger
\, \vec{a}_{2}^{\, [\mu\nu]\alpha} \, \vec{\tau} \, \tilde{K}%
^{\ast}_{\mu\nu} \, \ + \ \mathrm{h.c.} \, ) \, \\[3mm]
&+& \frac{1}{\sqrt{2}} \, \tilde{K}_{\mu\nu}^{\ast \, \dagger} [ g_{K_2^\ast
\pi K^\ast} \,\, i \, \partial_\alpha \vec{\pi} \, \vec{\tau} \, + \,
g_{K_2^\ast \eta K^\ast} \,\, i \, \partial_\alpha \eta \, ] \, K_{2}^{\ast
\, [\mu\nu]\alpha} \, \ + \ \mathrm{h.c.}  \nonumber \\[3mm]
&+& \frac{1}{\sqrt{2}} \, K_{2}^{\ast \, [\mu\nu]\alpha \, \dagger} \, [
g_{K_2^\ast K \rho} \,\, \tilde{\vec{\rho}}_{\mu\nu} \vec{\tau} \, + \,
g_{K_2^\ast K \omega} \,\, \tilde{\omega}_{\mu\nu} ] \, i \, \partial_\alpha
K \, \ + \ \mathrm{h.c.} \,.  \nonumber
\end{eqnarray}

\vspace*{.5cm}

\begin{center}
\textbf{Table 4.} $T\rightarrow PV$ coefficients.

\vspace*{0.5cm}

\begin{tabular}{|l|l|l|l|}
\hline
\hspace*{.6cm}$T$ & \hspace*{.3cm} $PV$ & $\alpha_{TPV}$ & $\hspace*{.6cm}
\tilde g_{TPV}$ \\ \hline
$f_{2}^{\prime }(1525)$ & \hspace*{.3cm} $K K^{\ast}$ & \hspace*{.3cm}$4$ & 
\hspace*{.2cm} $\sqrt{3} \, \cos\theta_{T}$ \\ \hline
$a_{2}(1320)$ & \hspace*{.3cm} $\pi \rho $ & \hspace*{.3cm}$2$ & \hspace*{%
.7cm} $2$ \\ \hline
$a_{2}(1320)$ & \hspace*{.2cm} $KK^{\ast }$ & \hspace*{.15cm} $2$ & \hspace*{%
.7cm} $1$ \\ \hline
$K_{2}(1430)$ & \hspace*{.35cm}$\pi K^{\ast}$ & \hspace*{.3cm}$3$ & \hspace*{%
.8cm}$1$ \\ \hline
$K_{2}(1430)$ & \hspace*{.25cm} $\eta K^{\ast}$ & \hspace*{.3cm}$2$ & 
\hspace*{.2cm} $\sqrt{3} \, \cos\theta_{P}$ \\ \hline
$K_{2}(1430)$ & \hspace*{.2cm} $K \rho $ & \hspace*{.3cm}$3$ & \hspace*{.5cm}
$1$ \\ \hline
$K_{2}(1430)$ & \hspace*{.2cm} $K\omega $ & \hspace*{.3cm}$2$ & \hspace*{.2cm%
} $\sqrt{3}\sin\theta_{V}$ \\ \hline
\end{tabular}
\end{center}

\vspace*{1cm}

The $T \to \gamma\gamma$ decay width reads (see also Refs.~\cite%
{Suzuki:1993zs,Bellucci:1994eb}): 
\begin{equation}
\Gamma _{T\gamma \gamma} \, = \, \frac{\alpha}{20} \ M_{T}^{3} \
g_{T\gamma\gamma}^{2} \,,
\end{equation}
where $\alpha = e^2/4\pi = 1/137$. The couplings $\tilde g_{T\gamma\gamma} =
g_{T\gamma\gamma}/c_{T\gamma\gamma}$ are given in Table 5.

\vspace*{.5cm}

\begin{center}
\textbf{Table 5.} $T\rightarrow \gamma \gamma $ coefficients.

\vspace*{0.5cm}

\begin{tabular}{|l|l|}
\hline
\hspace*{.6cm}$T$ & \hspace*{1.5cm}$\tilde g_{T\gamma \gamma }$ \\ \hline
$f_{2}(1270)$ & $\frac{4}{3 \sqrt{6}} \, ( 2 \sqrt{2} \cos\theta_T +
\sin\theta_T )$ \\ \hline
$f_{2}^{\prime }(1525)$ & $\frac{4}{3 \sqrt{6}} \, ( - 2 \sqrt{2}
\sin\theta_T + \cos\theta_T )$ \\ \hline
$a_{2}^{0}(1320)$ & \hspace*{1.5cm}$\frac{4}{3\sqrt{2}}$ \\ \hline
\end{tabular}
\end{center}

The $T \to P \gamma$ decay width reads: 
\begin{eqnarray}
\Gamma _{TP\gamma} \, = \, \frac{2}{5} \ \alpha \ P_{TP\gamma}^5 \
g_{TP\gamma}^2 \,,
\end{eqnarray}
where $P_{TP\gamma} = (M_T^2 - M_P^2)/2M_T$ is the three-momentum of final
states, $g_{TP\gamma}$ is the coupling constant from the tree-level
Lagrangian 
\begin{eqnarray}
\mathcal{L}_{TP\gamma} \, = \, \tilde F_{\mu\nu} \, \biggl\{ g_{a_2 \pi
\gamma} \, a_{2}^{- \, [\mu\nu]\alpha} \, i \, \partial_\alpha \pi^+ \ + \
g_{K_2^\ast K \gamma} \, K_{2}^{\ast - \, [\mu\nu]\alpha} \, i \,
\partial_\alpha K^+ \biggr\} \ + \ \mathrm{h.c.} \,.
\end{eqnarray}
The factors $\alpha_{TP\gamma}$ and scaled couplings $\tilde g_{TP\gamma} =
g_{TP\gamma}/c_{TP\gamma}$ are given in Table 6.

\vspace*{0.5cm}

\begin{center}
\textbf{Table 6.} $T\rightarrow P\gamma $ coefficients.

\vspace*{0.5cm}

\begin{tabular}{|l|l|l|}
\hline
\hspace*{.6cm}$T$ & $P$ & $\tilde g_{TP\gamma }$ \\ \hline
$a_{2}^{\pm }(1320)$ & $\pi ^{\pm }$ & \hspace*{.2cm} $1$ \\ \hline
$K_{2}^{\ast \, \pm }(1430)$ & $K^{\pm }$ & \hspace*{.2cm} $1$ \\ \hline
\end{tabular}
\end{center}

The decay width of the tensor-glueball $G$ into $P_1P_2$ pair reads 
\begin{eqnarray}
\Gamma _{GP_{1}P_{2}} \, = \, \alpha_{GP_1P_2} \, \frac{P_{GP_1P_2}^5}{60 \,
\pi \, M_G^2} \, g_{GP_1P_2}^2
\end{eqnarray}
where the corresponding parameters $\alpha_{GP_1P_2}$ and the coupling
constants $\tilde g_{GP_1P_2} = g_{GP_1P_2}/c_{GP_1P_2}$ are given in
Table~7. The couplings $g_{GP_1P_2}$ arise from the tree-level Lagrangian 
\begin{eqnarray}  \label{L_GPP}
\mathcal{L}_{GP_1P_2} &=& G^{\mu\nu} \, \biggl\{ g_{G\pi\pi} \, \partial_\mu%
\vec{\pi} \, \partial_\nu\vec{\pi} \, + \, 2 \, g_{G \overline K K} \,
\partial_\mu K^\dagger \, \partial_\nu K \, + \, g_{G\eta\eta} \,
\partial_\mu\eta \, \partial_\nu\eta \, + \, g_{G\eta^\prime\eta^\prime} \,
\partial_\mu\eta^\prime \, \partial_\nu\eta^\prime \biggr\} \,
\end{eqnarray}
are, as expected, equal to each other for all the modes, apart from the
forbidden $\eta \eta ^{\prime }$ one.

\vspace*{0.5cm}

\begin{center}
\textbf{Table 7.} $G\rightarrow P_1P_2$ coefficients.

\vspace*{0.5cm}

\begin{tabular}{|l|l|l|}
\hline
$P_{1}P_{2}$ & $\alpha_{GP_1P_2}$ & $\tilde g_{GP_1P_2}$ \\ \hline
$\pi \pi $ & \hspace*{.3cm} $6$ & \hspace*{.3cm} $1$ \\ \hline
$\overline{K}K$ & \hspace*{.3cm} $8$ & \hspace*{.3cm} $1$ \\ \hline
$\eta \eta $ & \hspace*{.3cm} $2$ & \hspace*{.3cm} $1$ \\ \hline
$\eta^{\prime }\eta^{\prime }$ & \hspace*{.3cm} $2$ & \hspace*{.3cm} $1$ \\ 
\hline
\end{tabular}
\bigskip\ 
\end{center}

The decay rate of the glueball into two vectors following from the
non-derivative coupling (see Eqs.~(\ref{Leff_G}) and (\ref{U3_breaking})) 
reads: 
\begin{eqnarray}  \label{Gamma_GVV}
\Gamma_{GV_{1}V_{2}} \, = \, \alpha_{GV_{1}V_{2}}\frac{P_{GV_1V_2}}{8
\,\pi\,M_{G}^{2}} \, F_{GV_1V_2} \, g_{GV_1V_2}^2
\end{eqnarray}
where 
\begin{eqnarray}
F_{GV_1V_2} \, = \, 1+\frac{P_{GV_1V_2}^2}{3} \, \left( \frac{1}{M_{V_1}^2}
\, + \, \frac{1}{M_{V_2}^2} \right) +\frac{2}{15} \, \frac{P_{GV_1V_2}^4}{%
M_{V_1}^2 \, M_{V_2}^2} \,.
\end{eqnarray}
In the case of derivative coupling, the function $F_{GV_1V_2}$ changes (see
Ref.~\cite{Suzuki:1993zs}) but the factors $\alpha_{GV_1V_2}$ and $%
g_{GV_1V_2}$ do not. As it was mentioned before we consider two specific
scenarios: i) the case of $U(3)$ symmetry when the octet $(c_{GVV}^8)$ and
singlet $(c_{GVV}^0)$ couplings degenerate $c_{GVV}^8 = c_{GVV}^0 = c_{GVV}$%
; ii) the case of the broken $U(3)$ symmetry to $SU(3)$ one. For convenience
we put $c_{GVV}^8 = c_{GVV}$ and introduce the breaking parameter $y_{GVV} =
c_{GVV}^0/c_{GVV}^8$ which is equal to one at large $N_c$ limit (or in case
of the $U(3)$ invariance). The couplings $g_{GV_1V_2}$ arise from the
three-level Lagrangian 
\begin{eqnarray}  \label{L_GVV}
\mathcal{L}_{GV_1V_2} &=& G^{\mu\nu} \ \biggl\{ \, g_{G\rho\rho} \, \vec{\rho%
}_\mu \, \vec{\rho}_\nu \, + \, 2 \, g_{G \overline K^\ast K^\ast} \,
K^{\ast \dagger}_\mu \, K^{\ast}_\nu \, + \, g_{G\omega\omega} \, \omega_\mu
\, \omega_\nu \, + \, g_{G\omega\phi} \, \omega_\mu \, \phi_\nu \, + \,
g_{G\phi\phi} \, \phi_\mu \, \phi_\nu \, \biggr\} \,.
\end{eqnarray}
The parameters occurring in Eq.~(\ref{Gamma_GVV}) are given in Table 8.
Again we rescale $g_{GV_1V_2}$ by the coupling $c_{GVV}^8$ as $\tilde
g_{GV_1V_2} = g_{GV_1V_2}/c_{GVV}^8$. In the last two columns of Table 8 we
present the results for the couplings $\tilde g_{GV_1V_2}$ for the $U(3)$
symmetric case and for the case of the broken $U(3)$~symmetry.

\vspace*{0.5cm}

\begin{center}
\textbf{Table 8: }$G\rightarrow VV$ coefficients

\vspace*{0.5cm}

\begin{tabular}{|l|l|l|l|}
\hline
&  & \hspace*{0.5cm} $\tilde{g}_{GV_{1}V_{2}}$ & \hspace*{1.4cm} $\tilde{g}%
_{GV_{1}V_{2}}$ \\ 
\hspace*{0.5cm}$V_{1}V_{2}$ & $\alpha _{GV_{1}V_{2}}$ & \hspace*{0.025cm} [ $%
y_{GVV}=1$ ] & \hspace*{0.85cm} [ any $y_{GVV}$ ] \\ \hline
\hspace*{0.5cm}$\rho \rho $ & \hspace*{0.3cm} $6$ & \hspace*{0.9cm} $1$ & 
\hspace*{1.7cm} $1$ \\ \hline
\hspace*{0.35cm}$\overline{K}^{\ast }K^{\ast }$ & \hspace*{0.3cm} $8$ & 
\hspace*{1cm}$1$ & \hspace*{1.7cm} $1$ \\ \hline
\hspace*{0.5cm}$\omega \omega $ & \hspace*{0.3cm} $2$ & \hspace*{0.9cm} $1$
& \hspace*{0.3cm}$\cos ^{2}\theta _{V}\,y_{GVV}\,+\,\sin ^{2}\theta _{V}$ \\ 
\hline
\hspace*{0.5cm}$\omega \phi $ & \hspace*{0.3cm} $1$ & \hspace*{1cm}$0$ & 
\hspace*{0.2cm} $\sin 2\theta _{V}\,(1-y_{GVV})$ \\ \hline
\hspace*{0.5cm}$\phi \phi $ & \hspace*{0.3cm} $2$ & \hspace*{0.9cm} $1$ & 
\hspace*{0.3cm}$\sin ^{2}\theta _{V}\,y_{GVV}\,+\,\cos ^{2}\theta _{V}$ \\ 
\hline
\end{tabular}
\bigskip

\bigskip 
\end{center}

\end{document}